\documentclass[letterpaper, 10 pt, conference]{ieeeconf}  

\IEEEoverridecommandlockouts                              
\usepackage{amsmath} 
\usepackage{amssymb}  

\usepackage{url}
\usepackage[ruled, vlined, linesnumbered]{algorithm2e}
\usepackage{verbatim} 
\usepackage{soul, color}
\usepackage{lmodern}
\usepackage{fancyhdr}
\usepackage[utf8]{inputenc}
\usepackage{fourier} 
\usepackage{array}
\usepackage{makecell}
\usepackage{scicite}
\usepackage{float}
\usepackage{upgreek}
\usepackage{graphicx}
\usepackage{subfigure}

\UseRawInputEncoding

\SetNlSty{large}{}{:}

\makeatletter

\newcommand{\Rom}[1]{\expandafter\@slowromancap\romannumeral #1@}
\makeatother

\title{\LARGE
Experimental realisation of a universal inverse-design magnonic device
}

\author{Noura Zenbaa,$^{1,2}$ Claas Abert,$^{1}$ Fabian Majcen,$^{1}$ Michael Kerber,$^{1}$ \\ Rostyslav O. Serha,$^{1,2}$ Sebastian Knauer,$^{1}$ Qi Wang,$^{3}$ Thomas Schrefl,$^{4}$ \\Dieter Suess,$^{1}$ Andrii V. Chumak$^{1\ast}$\\
\\
\normalsize{$^{1}$University of Vienna, Faculty of Physics, Vienna 1090, Austria}\\
\normalsize{$^{2}$University of Vienna, Vienna Doctoral School in Physics, Vienna 1090, Austria}\\
\normalsize{$^{3}$Huazhong University of Science and Technology, School of Physics, Wuhan 430074, China}\\
\normalsize{$^{4}$Donau-Universit\"at Krems, Center for Modelling and Simulation, Wiener Neustadt 2700,
Austria}
\\
\normalsize{$^\ast$To whom correspondence should be addressed; E-mail:  andrii.chumak@univie.ac.at}
}

\begin{document}

\maketitle
\thispagestyle{plain}
\pagestyle{plain}

\begin{abstract}

In the field of magnonics, which uses magnons, the quanta of spin waves, for energy-efficient data processing, significant progress has been made leveraging the capabilities of the inverse design concept. This approach involves defining a desired functionality and employing a feedback-loop algorithm to optimise the device design. In this study, we present the first experimental demonstration of a reconfigurable, lithography-free, and simulation-free inverse-design device capable of implementing various RF components. The device features a square array of independent direct current loops that generate a complex reconfigurable magnetic medium atop a Yttrium-Iron-Garnet (YIG) rectangular film for data processing in the gigahertz range. Showcasing its versatility, the device addresses inverse problems using two algorithms to create RF notch filters and demultiplexers. Additionally, the device holds promise for binary, reservoir, and neuromorphic computing applications.\\

\end{abstract}


\section{INTRODUCTION}\label{sec1}

The fields of electronics and telecommunications ongoing research greatly revolves around frameworks like the Internet of Things (IoT), 5G and 6G technologies \cite{noauthor_5g_nodate}. These frameworks require RF devices that support high-speed and high-capacity wireless communication, which are vastly used in our mobile phones, autonomous vehicles, smart cities, and advanced healthcare monitoring systems \cite{giribaldi_compact_2024}. Therefore there is a continuous need to design, test, and optimise new advanced quickly-reconfigurable RF components that operate in a wide range of frequencies and bandwidths with low energy consumption. Magnons, the quanta of spin waves, are great candidates to use in realising RF components \cite{dieny_opportunities_2020}. Spin waves are the collective precessional motion of magnetic moments in a magnetic material that propagates as a wave~\cite{gurevich_magnetization_1996, stancil_spin_2009}, which do not involve any movements of particles and therefore is an energy-efficient way of data transfer and processing~\cite{v_v_kruglyak_magnonics_2010, barman_2021_2021, chumak_advances_2022}. Spin waves operate in the range of sub-GHz up to THz frequencies covering the 5G and the 6G regimes~\cite{barman_2021_2021,wu_high-performance_2017}. In addition, the wide range of nonlinear and nonreciprocal spin-wave physical phenomena opens up unique opportunities for the realisation of RF devices ~\cite{barman_2021_2021, chumak_advances_2022}.

In recent years, several fully functioning magnonic devices have been successfully demonstrated~\cite{wang_nanoscale_2024}, including directional couplers serving as building blocks for all-magnonic circuits~\cite{wang_magnonic_2020}, magnon valves and transistors~\cite{wu_magnon_2018, chumak_magnon_2014}, spin-wave logic gates~\cite{mahmoud_introduction_2020} and neuromorphic computing elements~\cite{torrejon_neuromorphic_2017, papp_nanoscale_2017, bracher_analog_2018}. However, the design process for each requires specialized investigation that takes a long time, and usually, one device performs one single functionality. The inverse design approach, which was recently implemented in the field of magnonics and provided a great momentum, offers an excellent solution. It is a two-step process that involves dividing a design area into smaller elements arranged in an array and by properly tuning these elements using feedback loop-based optimisation, any predefined functionality can be achieved. Magnonic (de-)multiplexers, nonlinear switches, Y-circulators~\cite{wang_inverse-design_2021}, lenses~\cite{kiechle_experimental_2022}, and neuromorphic networks for vowel recognition~\cite{papp_nanoscale_2021} have been demonstrated showing the potential of the inverse design method. All the reported approaches, however, have major drawbacks, they require both time- and energy-consuming complex numerical computations. For complex systems, simulations of a single state take too long, in the range of minutes to hours while experiments are much faster, in the range of seconds down to sub-microseconds. Moreover, the experimental realisation of a simulated device is only possible in the case of a perfect match between the numerical simulations and the experiment. Reconfigurable magnonic devices realised by inverse design directly in the experiment are therefore in high demand for all types of magnon-based data processing including RF applications, binary and unconventional computing~\cite{wang_nanoscale_2024}.

\begin{figure*}[h]
\centering
\includegraphics[width=\textwidth]{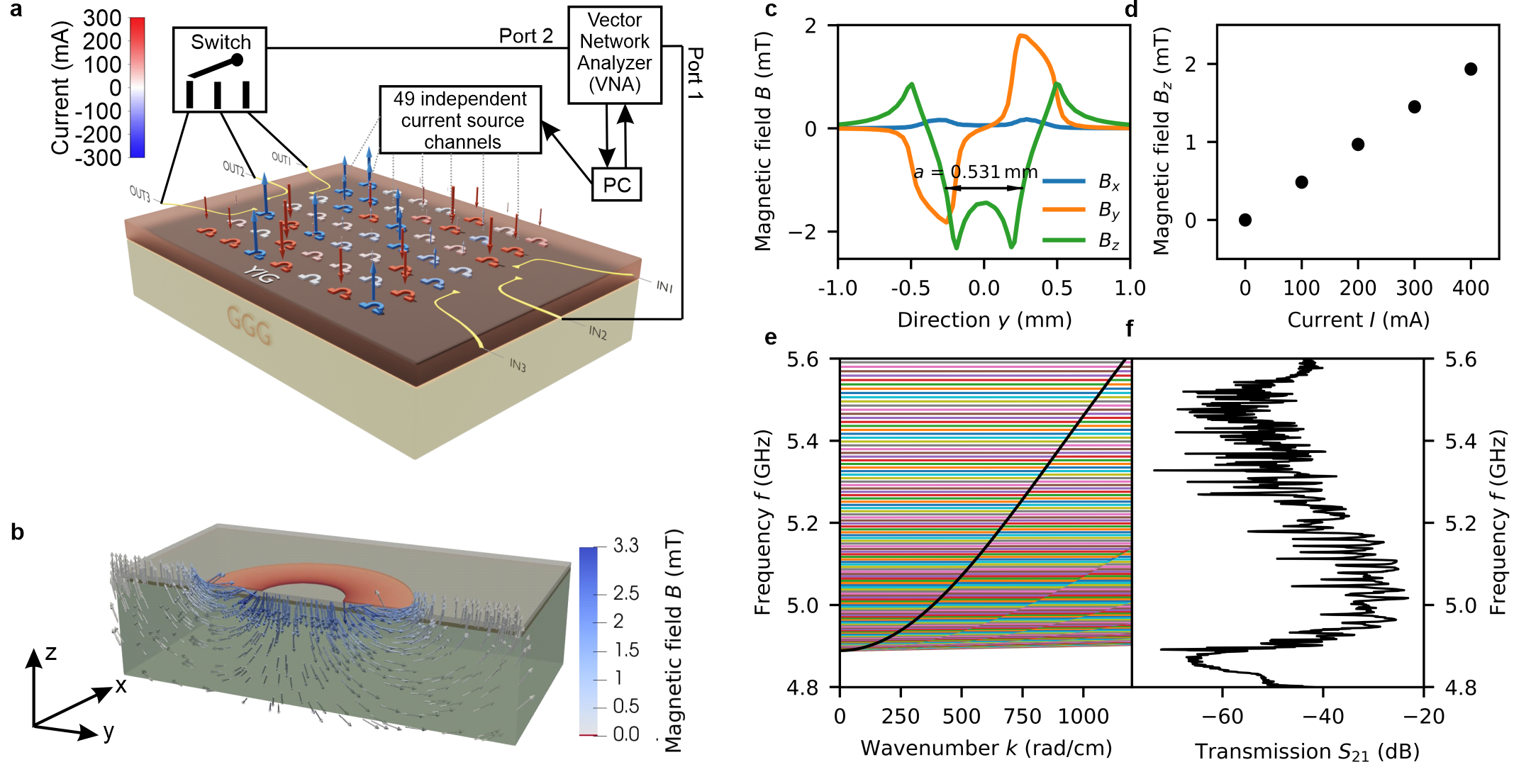}
\caption{{\bf Device design (a)} The magnonic inverse-design device based on a micrometer-thick YIG film, PC-controlled 49 reconfigurable omega-shaped current loops, three input and three output transducers, a VNA, independent current source channels, a switch and a magnet applying an out-of-plane bias field to excite isotropic FVMSWs. \textbf{(b)} The field generated by one omega loop through the YIG film simulation by magnum.pi. \textbf{(c)} Magnetic field magnitude generated by one loop $B$ as a function of direction \textit{y} defined at the center of omega loop at -300~mA applied current. \textbf{(d)} Out-of-plane magnetic field $B_z$ generated by one loop as a function of applied current $I$ at the center of the loop at the YIG surface. \textbf{(e)} The dispersion relation (frequency $f$ versus wavenumber $k$) of FVMSWs in the 18 $\upmu$m-thich YIG film~\cite{bozhko_unconventional_2020}, the bold curve is the fundamental mode and the horizontal modes correspond to PSSWs. \textbf{(f)} A reference transmission spectrum ($S^{\text{ref}}_{21}$ (dB) as a function of frequency) of the same YIG film between $I_{N2}$ and $O_{UT3}$.}\label{fig1}
\end{figure*}

Here, we report on the experimental realisation of a reconfigurable inverse-design device that operates with magnons in the GHz frequency range as a uniform platform for magnon-based data processing. It is based on a matrix of 7x7 direct current (DC) loops producing a static magnetic field in 2048 steps (offering 2048$^{\text{49}}$ $\approx$ 10$^{\text{162}} $ degrees of freedom, ~10$^{\text{87}}$ states were used in the measurements for practical reasons), to generate a complex and reconfigurable, fundamentally in the ns timescale~\cite{chumak_all-linear_2010}, magnetic field pattern -- see Fig.~\ref{fig1}a. The transmission of the spin wave through such a complex reconfigurable medium is measured experimentally after it has undergone multiple wave scattering, changing its direction, wavelength and phase due to linear and non-linear processes. The resulted transmission is processed by a specialised algorithm that defines the magnetic pattern, and the corresponding 49 current values, to be tested in the next iteration dependent on the user-defined functionality. The procedure is repeated until the objective function, which is a measure of performance in terms of the defined functionality, is maximised. The device is composed of multiple input and output spin-wave transducers to have access to a wide range of functionalities. Two different feedback-loop algorithms, Direct Search~(DS) optimisation, and a Genetic Algorithm~(GA), were successfully used to configure the field patterns to realise linear RF notch filter and demultiplexer. 

\section{Results}\label{sec2}
\subsection{Device design}\label{subsec2}
Our reconfigurable magnonic inverse-design device is based on a rectangular-shaped micron-thick Yttrium Iron Garnet~(YIG) film, a material allowing for long-range spin-wave transport~\cite{dubs_sub-micrometer_2017}. The YIG sample is placed below a printed circuit board (PCB) comprising of a 7$\times$7 omega-shaped DC loop array (see Fig.~\ref{fig1}a). Three input and three output spin-wave transducers are incorporated to access a wide range of device functionalities, but only some are used in the experiments reported here. The microstrip transducers allow for direct coupling between the magnetisation precession in the YIG film and the driving Oersted field of the microwave current in the wide wavelength range from millimeters down to around 50\,$\upmu$m~\cite{serga_yig_2010}. The measurement setup shown in Fig.~\ref{fig1}a uses an electromagnet to apply a biasing magnetic field perpendicular to the YIG surface to excite isotropic Forward Volume Magnetostatic Spin Waves~(FVMSWs)~\cite{gurevich_magnetization_1996, stancil_spin_2009}. The bias field was kept at 350\,mT, allowing operation with the propagating spin waves in the frequency range of 4.9 to 5.5\,GHz. A Vector Network Analyzer~(VNA) combined with a mechanical microwave switch is used to send/receive microwave signals to/from the different spin-wave transducers. The omega-shaped loops are connected to 49 independent current source channels, with a current range of $\pm$400\,mA. Each omega-shaped loop generates an Oersted field in the YIG film either parallel or antiparallel to the external bias field -- see the results of numerical simulations for a single current loop using magnum.pi~\cite{abert_magnumfe_2013,abert_micromagnetics_2019,schrefl_numerical_2007} in Fig.~\ref{fig1}b. The field spatial distribution across the y-direction (at the center of the omega loop exactly on the YIG surface) is displayed in Fig.~\ref{fig1}c and its amplitude as a function of current is shown in Fig.~\ref{fig1}d. A field of around $\pm$2\,mT is generated for the $\pm$400\,mA applied current. A spin wave, while propagating into the inhomogeneous magnetic field regions created by the omega loops, shifts its dispersion to higher or lower frequencies depending on the current polarity. Consequently, the spin waves of the same global frequency will either propagate with phase accumulations or will experience scattering off the inhomogeneous field regions, interfering in a non-intuitive way. In addition, the areas of the interference pattern with higher spin-wave density will undergo more pronounced nonlinear effects~\cite{wang_magnonic_2020, wang_deeply_2023}, while the spin waves of small amplitude will remain in the linear regime~\cite{papp_nanoscale_2021}. 

The reference transmission $S_{21}$ signal (measured between VNA ports 1 and 2) of the propagating spin waves, taken at zero currents applied to the omega loops, is plotted in Fig.~\ref{fig1}f. We measure insertion losses of about 25\,dB, which is related to the spin-wave loss during propagation over the distance of 2.2\,cm and to the relatively low efficiency of spin-wave excitation and detection by the transducers~\cite{connelly_efficient_2021}. The low-transmission regions (dips) in the spectrum arise from the hybridisation of the FVMSW mode with Perpendicular Standing Spin Waves~(PSSWs)~\cite{kalinikos_theory_1986,serga_parametrically_2007, vilsmeier_spatial_2024} (see dispersion in  Fig.~\ref{fig1}e) and do not play any positive or negative role in the operation of the inverse-design device. The comparison of the fundamental FVMSW mode (bold) dispersion curve with the transmission $S_{21}$ spectrum shows that the spin-wave wavelength working region spans from about 10\,mm down to 50\,$\upmu$m. This is of importance as the ability of the inverse-design prototype device to operate with waves of different wavelengths was tested and is presented below. 

\begin{figure*}[h]
\centering
\includegraphics[width=\textwidth]{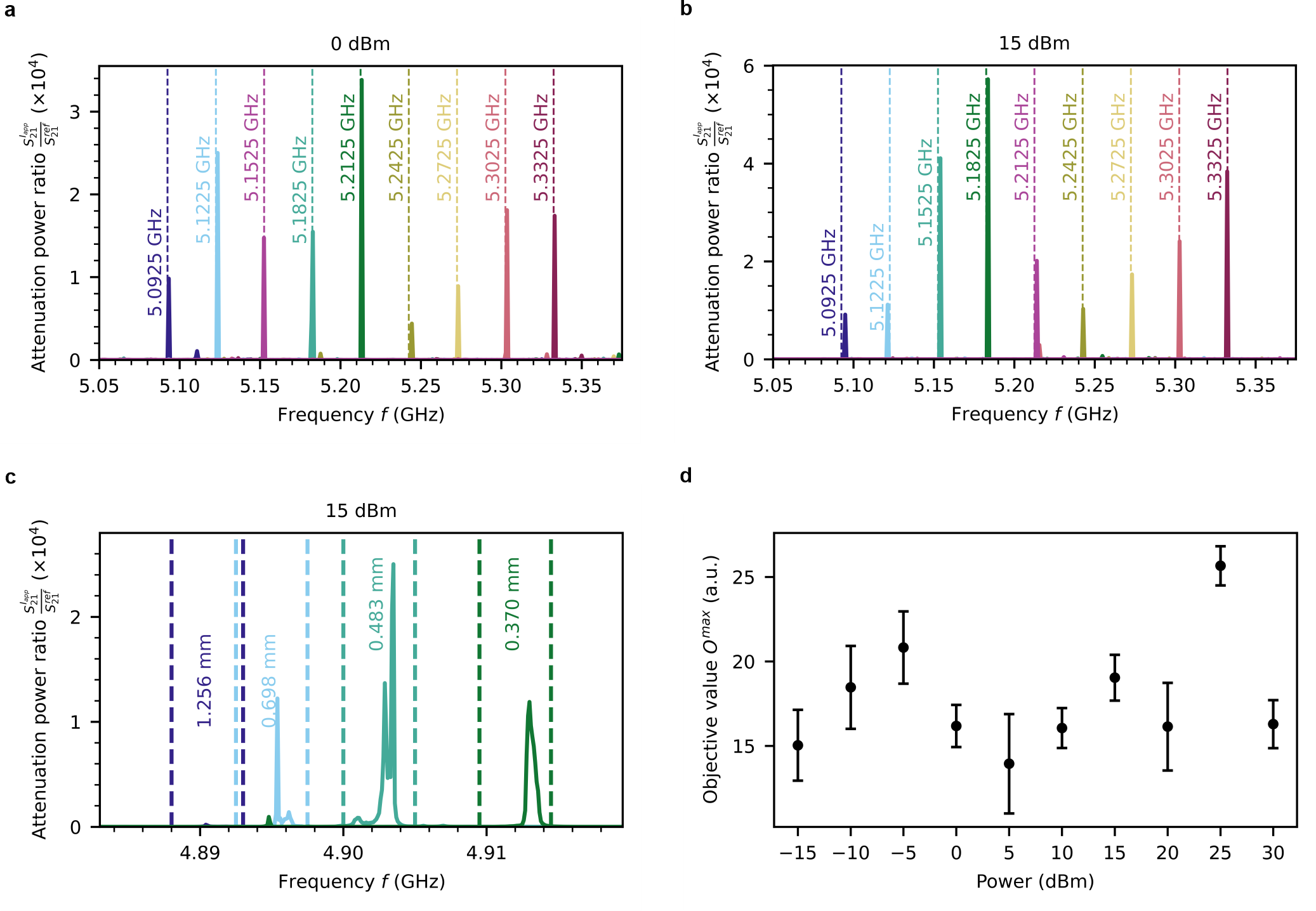}
\caption{{\bf Notch filter} between input $I_{N2}$ and $O_{UT3}$. Attenuation power ratio $(S^{\text{I}_{\text{app}}}_{21,j})/(S^{\text{ref}}_{21})$ as a function of frequency of nine 5\,MHz BW notch filters of different center frequencies in the \textbf{(a)} linear regime at 0\,dBm input power, and \textbf{(b)} weakly non-linear regime at 15\,dBm input power. \textbf{(c)} Notch filters characteristics at frequencies corresponding to the long wavelength range covering from 370\,$\upmu$m up to $\approx$ 1.3\,mm at 15\,dBm. \textbf{(d)} A power study showing objective value as a function of input power applied for the functionality of a 5\,MHz BW notch filter of 5.1775\,GHz center frequency.}\label{fig2}
\end{figure*}

\subsection{Demonstrator 1: Notch filter}\label{subsec4}
A notch filter, also known as a band stop filter or a band reject filter, is used abundantly in different RF technologies to suppress undesirable signals at specific frequency ranges from the original signal, like noise. To demonstrate that the presented universal inverse-design device can act as a notch filter, we use only one input spin-wave transducer as a filter input and one output transducer as a filter output. The bandwidth (BW) of the filter was fixed to 5 MHz and its operation at two applied microwave powers of 0\,dBm and 15\,dBm (linear and weakly-nonlinear regimes) were tested. The same device was used to realise nine notch filters of different center frequencies shown in Figures~\ref{fig2}a and \ref{fig2}b. 

The first step in inverse design is to define the, so-called, objective function to be maximised by the optimisation algorithm that tunes the currents applied to the omega loops. The objective function of a notch filter is defined as follows: 

\begin{equation}
\centering
 O_l^{\text{GA}} = (S^{\text{ref}}_{21}-S^{\text{I}_{\text{app}}}_{21,l}){_{\text{filter\ BW}}}-\left|S^{\text{ref}}_{21}-S^{\text{I}_{\text{app}}}_{21,l}\right|{_{\text{non filter BW}}},\label{obj_fn_pygad}
\end{equation}

where $S^{\text{ref}}_{21}$ is the transmission parameter from port 1 $\rightarrow$ port 2 when zero currents are applied to the DC loops, $S^{\text{I}_{\text{app}}}_{21,l}$ is the transmission parameter from port 1 $\rightarrow$ port 2 when the currents of configuration $l$ are applied. The losses within the filter bandwidth (BW) are maximized to optimize the objective function while minimizing any changes occurring outside the filter bandwidth. The measuring procedure starts with recording one reference signal $S^{\text{ref}}_{21}(f)$ that is used throughout the whole optimisation process. Afterwards an algorithm (see Sec.~\ref{sec3} below, GA optimisation is used to realize the notch filter functionality) defines current values for all omega loops, commands the current sources to apply the current configuration defined and the $S^{\text{I}_{\text{app}}}_{21,l}(f)$ spectrum ($l=1$ in this case) is measured automatically again. As a next step, the algorithm uses Eqn.~\ref{obj_fn_pygad} to calculate the objective function after the first iteration, and defines the next current configuration. The optimisation process is an iterative approach that stops only when the desired objective value is achieved or when it has completed the defined maximum number of iterations.

The original spin-wave transmission spectrum after the optimisation has finished, is shown in Fig.~\ref{fig4}b and \ref{fig5}b in the Methods section~\ref{sec3}. Signal transmission within the bandwidth, with values ranging from -65 to -85 dB, was achieved.
In the following, the performance of the notch filters is shown in terms of the attenuation power ratio, to focus on the effects induced solely by the reconfigurable medium. The attenuation power ratio $\frac{S^{\text{I}_{\text{app}}}_{21,l}}{S^{\text{ref}}_{21}}$ is calculated as follows:

\begin{equation}
     \centering
     \frac{S^{\text{I}_{\text{app}}}_{21,l}}{S^{\text{ref}}_{21}} = \frac{1}{10^{\frac{-\Delta S_{21}(dB)}{10}}},\label{eqn_atten}
\end{equation}

where $\Delta S_{21}(dB) = (S^{\text{I}_{\text{app}}}_{21,l}-S^{\text{ref}}_{21})$ is the difference between the transmission signal of iteration $l$ and the transmission of the reference signal at every frequency point. This ratio subtracts the reference signal from the final signal of the filter to show the induced attenuation by the reconfigurable medium only. 

The center frequencies considered were 30 MHz apart starting with center frequency $f_1 =$ 5.0925\,GHz (stop band starts at 5.09\,GHz and ends at 5.095\,GHz) up to center frequency $f_9 =$ 5.3325\,GHz. The average attenuation power ratio achieved is around 10$^\text{4}$ for all center frequencies and both applied powers. Such a very high suppression of spin-wave transmission of four orders of magnitude within the defined BW clearly demonstrates the high performance of the proof-of-concept device and the feasibility of the experimental realisation of the inverse-design approach in general. Additionally, the ease with which the center notch frequencies can be adjusted demonstrates that the inverse-design approach is highly suitable for creating reconfigurable RF devices. This approach allows for quick changes in functionality by simply modifying the current configuration, once optimisations for different functionalities have been performed.

One of the conceptual questions of inverse design is the ability to perform operations for different wavelengths. The results are shown in Figures~\ref{fig2}a and \ref{fig2}b cover a wavelength range of 70$\upmu$m up to 110$\upmu$m. This clearly demonstrates that the realised functionality is not bound to a particular wavelength, e.g. defined by the characteristic dimension of the field inhomogeneity or by the spatial distance between the loops. However, the wavelengths used in the notch filters presented were always smaller than the characteristic dimension of the field inhomogeneity of about 0.53\,mm -- see Fig.~\ref{fig1}c. Figure~\ref{fig2}c depicts notch filters covering the lower range of $k$, representing wavelengths between 0.37\,mm and 1.256\,mm (a wavenumber step of 40\,rad/cm was used). It is evident that even when the wavelength is larger than twice the characteristic dimension of the field inhomogeneity, the inverse-design approach successfully optimised a notch filter. However, the performance decreased significantly, with the attenuation power ratio dropping from $10^4$ down to 200. Therefore, the diffraction limit must be taken into account.

A power study was conducted for one center frequency to compare the device performance at different input powers -- see Fig.~\ref{fig2}d. The comparison was made using the same initial random population for all power values and limiting the optimisation run to 50 generations in GA per power only. The error bar represents the 5 different optimisation runs performed at each of the powers, while the black circle represents the average of these 5 runs. The study shows more oscillations in terms of objective function until 5~dBm, which is to be expected in the linear regime. However, starting at 10~dBm, the weakly non-linear regime, we can see a trend of increasing objective values except at 20~dBm, with a maximum reached at 25~dBm. This is explained by the additional phase accumulations due to the non-linear dispersion shift~\cite{wang_deeply_2023}, which can achieve higher suppression and help in finding the solution to the inverse-design problem at hand faster. At 30~dBm, we see a clear drop in objective value, which is a known behavior explained by triggering of the stochastic four-magnon scattering processes~\cite{chumak_magnon_2014}, disturbing the operation of the device. This behavior qualitatively agrees with the power study performed in~\cite{papp_nanoscale_2021} and demonstrates that deterministic non-linear processes are favored by inverse-design approaches.

\subsection{Demonstrator 2: Two-port magnonic frequency demultiplexer}\label{subsec5}
A demultiplexer is a circuit that has one input and multiple outputs which is used to send signal to more than one device. Demultiplexers are used widely in digital systems, computer networks, and RF communication systems for data routing, data transmission in synchronous systems and to be able to select one signal from a mutual stream of signals.

\begin{figure}[H]
\centering
\includegraphics{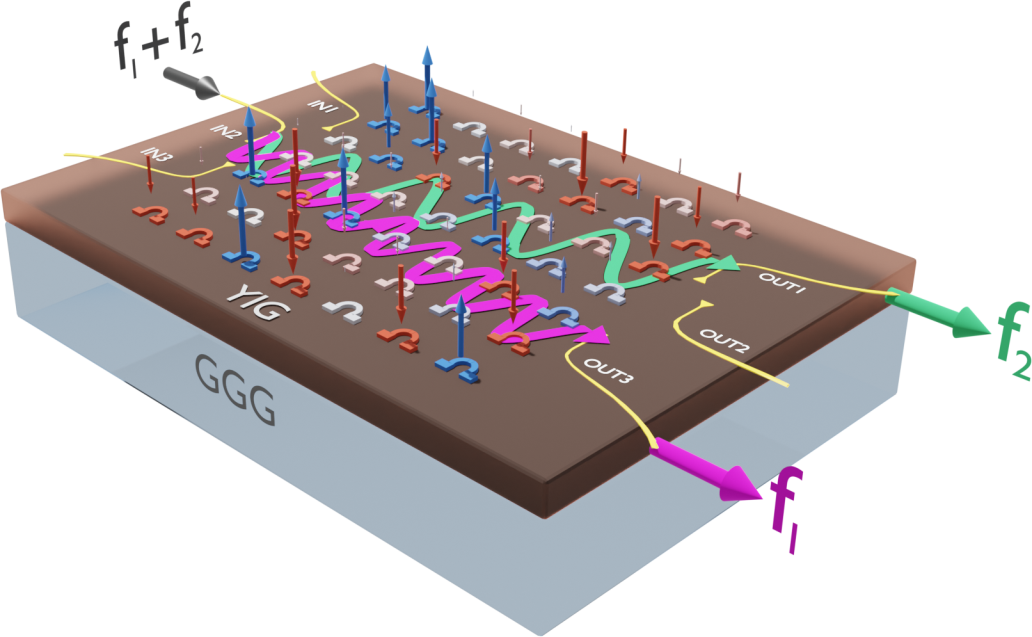}
\caption{{\bf Working principle of the inverse-designed two-port frequency demultiplexer.} Two frequencies $f_1$ and $f_2$ are excited at $I_{N2}$ transducer and guided through the design region such that $f_1$ is completely suppressed at $O_{UT1}$ transducer while being efficiently transmitted at $O_{UT3}$ transducer and vice-versa for $f_2$.}\label{demulti}
\end{figure}
We present a two-port frequency demultiplexer that utilizes one input $I_{N2}$ transducer and two output $O_{UT1}$ and $O_{UT3}$ transducers. Instead of using two fixed frequencies, the same VNA-based setup was used. The demultiplexer was realised by defining two 5\,MHz frequency ranges, with randomly selected center frequencies $f_1 =$ 5.1525\,GHz and $f_2 =$ 5.1825\,GHz. The spin waves of the swept frequencies are excited at $I_{N2}$. The optimisation algorithm is used to find the optimum current configuration to maximise $f_1$ transmitted signal at $O_{UT3}$ while minimising its transmission at $O_{UT1}$ and vice-versa for spin waves of $f_2$ -- see Fig.~\ref{demulti}. The optimisation algorithm used to achieve this two-port frequency demultiplexer is the Direct Search (DS) described in the Methods section~\ref{sec3} (see Fig.~\ref{fig5}a). The objective function was defined as the multiplication of the two output signals as follows: 

\begin{equation}
\centering
    O^{\text{DS}}_j = (S^{\text{ref}}_{21}-S^{\text{I}_{\text{app}}}_{21,j})_{f_1,O_{{UT}_1}}\times (S^{\text{ref}}_{21}-S^{\text{I}_{\text{app}}}_{21,j})_{f_2,O_{{UT}_3}}.\label{eqn4}
\end{equation}
\newline
\begin{figure*}[h]
\centering
\includegraphics[width=\textwidth]{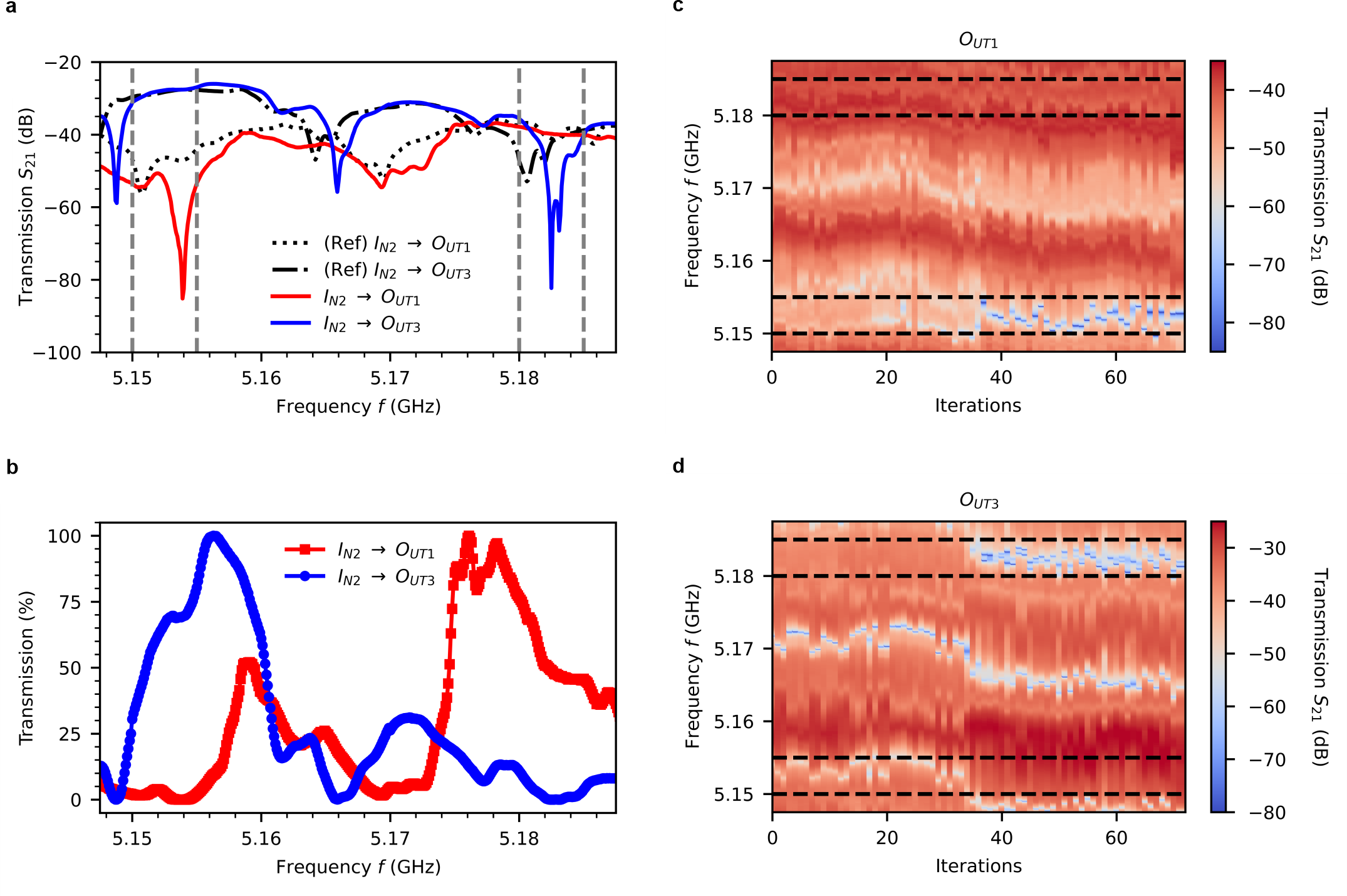}
\caption{{\bf Frequency de-multiplexer} \textbf{(a)} Transmission spectrum $S_{21}$ as a function of frequency of both outputs ($O_{UT1}$ and $O_{UT3}$) reference signals, and their respective transmission signals after the completion of the DS optimisation process. \textbf{(b)} Normalized transmission percentage of $O_{UT1}$ and $O_{UT3}$ as a function of frequency. One can see that the signal within the band 5.15 to 5.155\,GHz is efficiently detected at $O_{UT3}$ while being attenuated at $O_{UT1}$. On the other hand, the signal in the frequency band between 5.18 to 5.185\,GHz is detected at $O_{UT1}$ and suppressed at $O_{UT3}$. \textbf{(c)},\textbf{(d)} Color maps of the evolution of frequency rejection bands in the defined regions as a function of iteration number of \textbf{(c)} $O_{UT1}$, and \textbf{(d)} $O_{UT3}$.}\label{fig3}
\end{figure*}

The objective function achieves optimisation when it is maximised. It is designed to attenuate $f_1$ at $O_{UT1}$ and $f_2$ at $O_{UT3}$, effectively routing $f_1$ to $O_{UT3}$ and $f_2$ to $O_{UT1}$. The states where both components,$(S^{\text{ref}}_{21}-S^{\text{I}_{\text{app}}}_{21,j})_{f_1,O_{{UT}_1}}$ \& $(S^{\text{ref}}_{21}-S^{\text{I}_{\text{app}}}_{21,j})_{f_2,O_{{UT}_3}}$, exhibit evolution in the opposite direction (when both yield negative values) are taken into account and disregarded. Figure~\ref{fig3}a shows the transmission signal $S_{21}$ as a function of frequency, where VNA port 1 is connected to the input side of the device, and port 2 is connected to the mechanical switch that connects to $O_{UT1}$ and $O_{UT3}$. The measurement process begins by obtaining a reference transmission $S^{\text{ref}}_{21}$ signal (with zero currents applied) at $O_{UT1}$. Subsequently, the VNA port 2 is directed to $O_{UT3}$ through the mechanical switch, and the corresponding reference $S^{\text{ref}}_{21}$ signal is recorded. These reference signals serve as the basis for calculating all objective values once the current configurations instructed by the optimiser begin to be applied. The figure displays two reference signals corresponding to the two outputs, along with the demultiplexer's signal after the completion of the optimisation process. The optimisation process has successfully attenuated the transmission of $f_1$ and $f_2$ ranges at $O_{UT1}$ ($\approx$ -80~dB) and $O_{UT3}$  ($\approx$ -85~dB), respectively, per the defined objective. Simultaneously, it has maintained the transmission at the same level as their respective reference signals at the defined outputs ($f_1$ to $O_{UT3}$ and $f_2$ to $O_{UT1}$). The normalized transmission percentage shown in Fig.~\ref{fig3}b, clearly illustrates the extended attenuation window compared to the defined frequency ranges, approximately double the defined windows. This highlights the robustness of the demultiplexer. The 100\% transmission is a relative value of each independent output to itself. Figures~\ref{fig3}c and \ref{fig3}d show the color map of the transmission $S_{21}$ in dB when considering the frequency and the number of iterations for $O_{UT1}$ and $O_{UT3}$, respectively. The plots show the evolution of the attenuated signal in the defined frequency ranges (marked with dashed lines) of each output through the number of iterations. The emergence of distinct rejection bands for the two different outputs becomes evident after approximately 35\, iterations ($\approx$ 3.5 minutes), as indicated by the formation of blue regions (-85\,dB for $O_{UT1}$ and -80\,dB for $O_{UT3}$) within the specified attenuation frequency ranges. Conversely, those frequency ranges appear red when considering the opposing output, indicating high transmission.

The presented results clearly demonstrate the high performance of the proposed universal inverse-design device as a frequency demultiplexer. As has been shown earlier, the operational frequencies can be easily changed by training the device using a different objective function.

\section{Methods}\label{sec3}
\subsection{Sample and experimental setup}
The sample used in an 18-$\upmu$m-thick Yttrium Iron Garnet~(YIG) rectangular film (24$\times$17.5)\,mm$^2$ grown on a 500\,$\upmu$m Gadolinium Gallium Garnet~(GGG) by Liquid Phase Epitaxy (LPE)~\cite{dubs_sub-micrometer_2017}. LPE is a technique that ensures lattice-matched crystalline growth of YIG on GGG to minimise spin-wave damping.

The microwave transducers used are copper microstrip transmission lines printed on a duroid substrate that are matched to 50\,$\Omega$ impedance at their base and have width that decreases linearly down to 50\,$\upmu$m to excite spin waves in a wide range of wavenumber, ranging from 3.55\,rad/cm up to 0.111\,rad/$\upmu$m~\cite{serga_yig_2010}. The distance between input and output transducers was kept at about 2.2\,cm.

The PCB carrying the current loops consists of four metallic layers, with all 49 omega loops printed on the top layer in direct contact with the YIG sample. Each omega loop is 1.1\,mm wide, and the center-to-center distance between the loops is 2\,mm. The design region, where the loops are printed, spans an area of (15$\times$15)$\text{mm}^2$. A heat insulator Teflon layer of 50\,$\upmu$m thickness was placed between the YIG sample and the PCB to (1) avoid the spin wave scattering off the metallic loops and (2) prevent parasitic heating from the loops to reach the YIG sample.

The PC-controlled multi-channel current source with feedback loops, developed by Elbatech Srl., is designed to apply currents in the range of $\pm$1\,A for each channel.

An electromagnet was used to apply an out-of-plane external field, maintained at 350\,mT by a specially developed magnet driver. The driver continuously measured and adjusted the field throughout the measurement process, ensuring an accuracy up to the third decimal point in mT. Water-cooled magnet poles were also used to dissipate any parasitic heat generated by the current-loops-carrying PCB.

The setup used a two-port VNA in combination with a mechanical switch to both excite and detect spin-wave signals at the input and output sides of the device. The VNA used can excite and detect signals in the range from 10\,MHz up to 20\,GHz.

Algorithms and measurements were all run using Python-supported libraries.

\begin{figure*}[h]
\centering
\includegraphics[width=\textwidth]{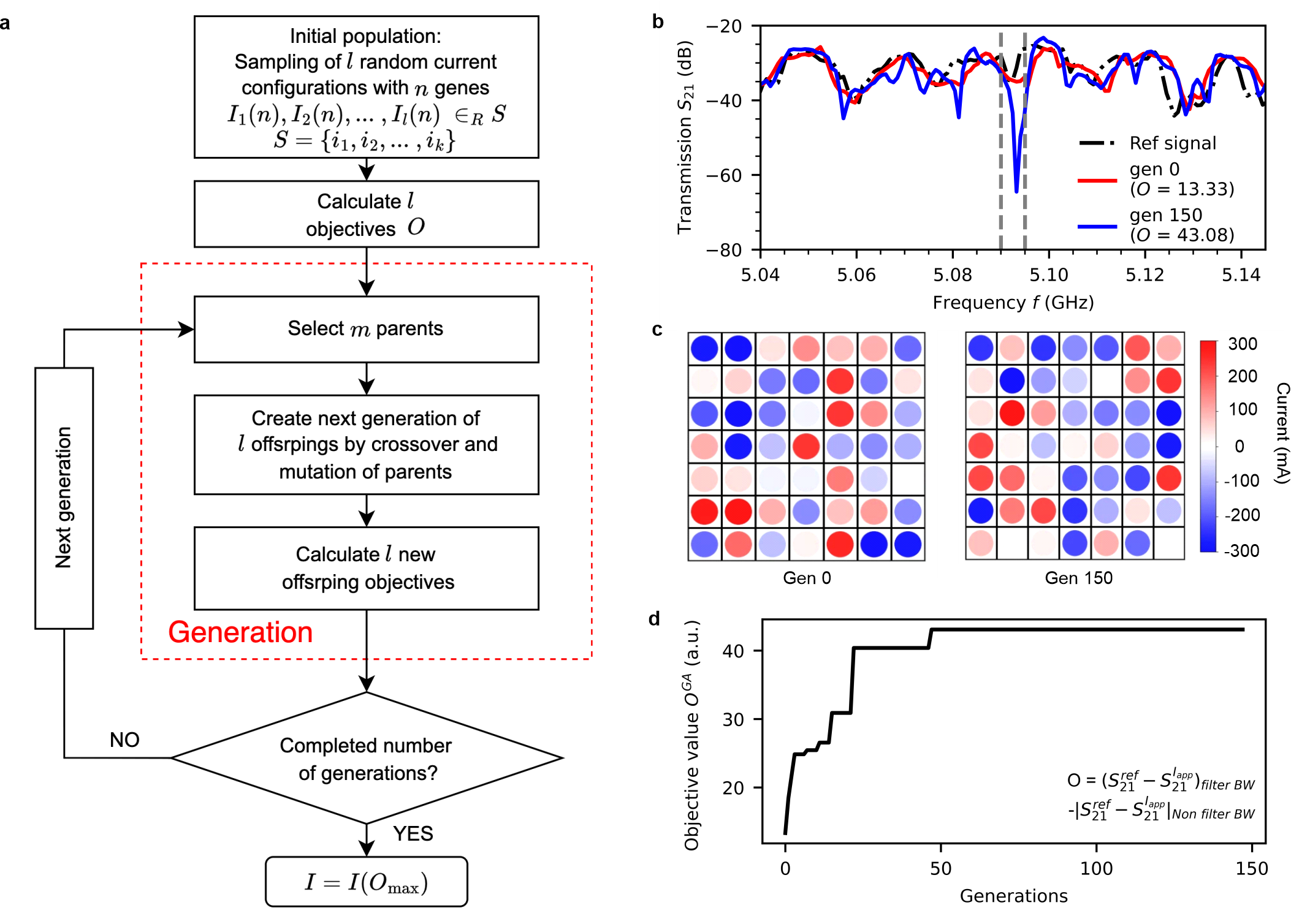}
\caption{{\bf The genetic algorithm} \textbf{(a)} The flow chart of the genetic algorithm process. \textbf{(b)} Transmission spectra of a reference signal, generations 0 and 150 signals of a 5\,MHz notch filter at a center frequency of 5.0925\,GHz. \textbf{(c)} The current distribution in the 49 loops of generation 0 and 150. \textbf{(d)} The evolution of objective value $O^{\text{GA}}$ as a function of the generation number (only improved states are shown). It shows the objective function defined for the notch filter shown in panel (b).}\label{fig4}
\end{figure*}

\subsection{Genetic Algorithm (GA)}
Two algorithms were used to test the performance of the inverse-design device. The first is a genetic algorithm representing a machine learning optimisation process. It relies on creating a new generation by the crossover of the best-performing parents and introducing mutations -- see Fig.~\ref{fig4}a~\cite{gad_pygad_2023}. The algorithm starts by creating a sample generation of $l$ random configurations with $n$ genes, $n$ is set to 49, corresponding to the 49 current values in the DC loops, and $l$ is assigned to the value of 50 in this concrete case, corresponding to the number of current loops. It provides the command for each current source channel to apply the corresponding current and then commands the VNA to measure $l$ spectra corresponding to the $l$ configurations. The transmission data is returned to the PC, and the optimiser calculates each configuration's objective value $O$. As a first example, we use this genetic algorithm to realise a notch filter functionality, where we only use one input and one output (between $I_{N2}$ and $O_{UT3}$), using the objective function defined in Eqn.~\ref{obj_fn_pygad}.

From the set of objectives calculated, the optimiser selects $m$ parents with the largest $O$, where $m$ is set to 10, and uses them to create the next generation of $l$ offspring. It calculates the $O$ of each offspring from the new generation, and iterates until it has completed the number of generations commanded. The evolution of objective value $O$ as a function of generation number is plotted in Fig.~\ref{fig4}d. An example of a 5\,MHz BW notch filter that is realised using the genetic algorithm is shown in Fig.~\ref{fig4}b. It compares the reference signal to the best offspring of generation 0 and the best in generation 150 for a notch filter with a center frequency of 5.0925\ GHz. The figure illustrates the signal's progression from generation 0 ($O =$ 13.33), where the maximum attenuation $(S^{\text{I}_{\text{app}}}_{21,l}-S^{\text{ref}}_{21})$ achieved is -8.3\,dB at $S^{\text{ref}}_{21} =$ -26.57\,dB compared to $S^{\text{I}{\text{app}}}_{21,0} =$ -34.89\,dB. By generation 150 ($O =$ 43.08), the maximum attenuation occurred at $S^{\text{ref}}_{21} =$ -31.3 dB compared to $S^{\text{I}{\text{app}}}_{21,150} =$ -64.5\,dB, resulting in -33.2\,dB attenuation, solely attributed to the applied current determined by the algorithm. The current distribution in the omega-shaped loops is shown for both generation 0 and generation 150 in Fig.~\ref{fig4}c with color coding in which red and blue correspond to opposite current polarities. 

The genetic algorithm, leveraging its adaptability and robustness, consistently yields optimal solutions in our experiments, showcasing its effectiveness in optimising and configuring complex systems.

\begin{figure*}[h]
\centering
\includegraphics[width=\textwidth]{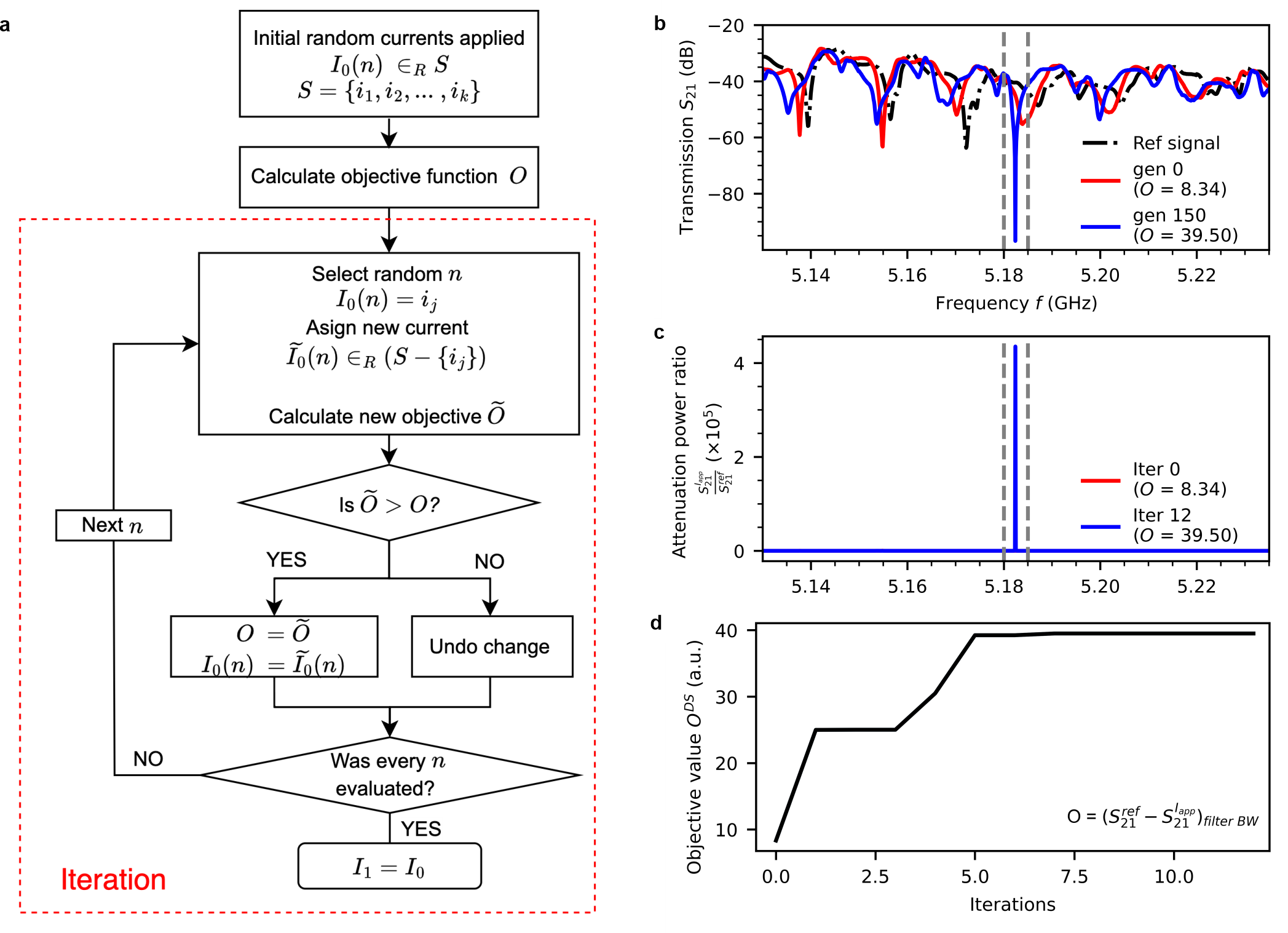}
\caption{{\bf The direct search algorithm} \textbf{(a)} The flow chart of the Direct Search (DS) algorithm. \textbf{(b)} transmission spectra of a reference signal, iterations 0 and 12 of a 5\,MHz notch filter at a center frequency of 5.1825\,GHz. \textbf{(c)} Attenuation power ratio $(S^{\text{I}_{\text{app}}}_{21,j})/(S^{\text{ref}}_{21})$ as a function of frequency of iteration 0 and 12. It shows attenuation of $\approx$ 4$\times$10$^\text{5}$ within the filter BW. \textbf{(d)} The evolution of objective value $O^{\text{DS}}$ as a function of the iteration number. It shows the objective function defined for the notch filter shown in panel (b).}
\label{fig5}
\end{figure*}

\subsection{Direct Search (DS) algorithm}
The second optimisation algorithm that has demonstrated high performance in the developed inverse-design device is a Direct Search~(DS) algorithm, which is very similar in concept to Direct Binary Search~(DBS) algorithms used in~\cite{wang_inverse-design_2021,shen_integrated-nanophotonics_2015}. It relies on a set of finite values defined by the user instead of the binary approach to find the optimum solution. It begins by creating an initial random current configuration $I_0(n)$ where $n$ is the variable that represents the number of the current loops in the array. The initial random configuration is a set of 49 current loops, each taking a random value from the set $S = (i_1, i_2, ...., i_k)$ -- see Fig.~\ref{fig5}a. It calculates the objective function $O$ of the initial state and then chooses a random loop $n$, checks its current value e.g. $i_j$, and chooses another current value from the set $S-\{i_j\}$. It calculates the new objective $O$ after the change, compares it to the original $O$, and decides to keep the previous current value or apply the new one based on which results in the largest objective value $O$. The optimiser continues over all 49 loops until each has been randomly changed once, one at a time. The first iteration is complete when it has gone over all the loops, progressing to the next state of $I_1(n)$. In Fig.~\ref{fig5}b, the resulting notch filter functionality at a center frequency of  5.1825\,GHz and a filter BW of 5\,MHz using DS is shown together with the reference signal and iteration 0. The plot illustrates the signal progress within the rejection band, beginning at iteration 0 with an initial value of $O =$ 8.34, corresponding to a maximum attenuation of -13.025\,dB at $S^{\text{I}{\text{app}}}_{21,0} =$ -55.24\,dB relative to $S^{\text{ref}}_{21} =$ -42.25 dB. By the final iteration ($O =$ 39.5), the attained attenuation reached -56.4\,dB, corresponding to $S^{\text{I}{\text{app}}}_{21,12} =$ -96.83\,dB compared to $S^{\text{ref}}_{21} =$ -40.45\,dB. The evolution of objective value $O$ over the number of iterations is shown in Fig.~\ref{fig5}d with the objective function defined as follows:
\begin{equation}
\centering
    O^{\text{DS}}_j = (S^{\text{ref}}_{21}-S^{\text{I}_{\text{app}}}_{21,j}){_{\text{filter\ BW}}}.\label{eqn2}
\end{equation}

In this case, the objective values were calculated by considering only the filter BW and not any changes that occurred outside the filter region since the measured changes outside the filter BW were insignificant.

Figure~\ref{fig5}c shows a 10$^\text{5}$ attenuation in the filter BW while maintaining the changes minimal outside the filter region. In many cases, this relatively straightforward DS algorithm has shown a high efficiency with as little as 12 iterations to find the optimised solution.

\section{Discussion}\label{sec4}
\subsection{The role of non-linearity}
The proposed magnonic device is an excellent proof-of-concept demonstrator of the general ability to solve an inverse problem experimentally. It is also a platform of choice for the further testing and development of different (including Artificial Intelligence (AI)-driven) optimisation algorithms that can be used for inverse-design tasks in 5G and 6G RF applications as well as in binary logic gates or neuromorphic data processing units. As it has been shown in photonics~\cite{molesky_inverse_2018,hughes_adjoint_2018} and magnonics~\cite{papp_nanoscale_2021}, the wave non-linearity, namely the ability of the wave to change the media by the increase in the wave amplitude, is important in inverse design. Spin waves are rich in versatile nonlinear phenomena inherent to their nature, eliminating the necessity for specialised nonlinear media. Roughly, one can split the phenomena into two types~\cite{gurevich_magnetization_1996}: controllable, which only slightly shifts the spin-wave dispersion curve inducing a phase accumulation of the wave~\cite{wang_reconfigurable_2018,wang_deeply_2023} as well as stochastic like three-magnon~\cite{ge_nanoscaled_2024} and four-magnon scattering~\cite{chumak_magnon_2014}. The stochastic processes are not desirable for the deterministic inverse-design approach, and the application of very large powers to the transducers deteriorates the operational characteristics of the devices, as has been shown in~\cite{papp_nanoscale_2021}. 

In this work, we have tested two regimes over a range of powers from -15 dBm to +30 dBm. In the linear regime (approximately for powers below +5 dBm), the spin-wave amplitude variation caused by the formation of a complex interference pattern is not significant enough to alter the medium; the spin waves exhibit the same dispersion across all the device regions. The system transitions to the nonlinear regime when the input power reaches +25 dBm. In this scenario, the large magnetization precession angle results in a reduction of the effective magnetization $M_{\text{eff}}$ of YIG \cite{wang_magnonic_2020,wang_deeply_2023}. Consequently, this change in magnetization shifts the spin-wave dispersion to higher frequencies (in the FVMSW geometry) and modifies the spin-wave wavelength and group velocity for spin waves of the same global frequency, which remains constant. Furthermore, this shift is spatially non-uniform and is defined by the spin-wave interference pattern, where localized regions with higher spin-wave amplitude experience a more pronounced dispersion shift. Thus, the nonlinearity introduces an additional non-uniform phase accumulation mechanism into the inverse design, enhancing its efficiency. Figure~\ref{fig2}d clearly shows that the higher objective value of around 25 was achieved in the nonlinear regime compared to around 15-20 in the linear regime.

In this manuscript, we report the realization of only linear functionalities. However, it is important to note that many tasks, such as neuromorphic computing or logic gates, inherently require the nonlinear regime and cannot be achieved in a linear system. The nonlinear functionalities of the proposed device are the subject of separate investigations.

\subsection{Operational speed}
If we analyze the speed with which the device can be reconfigured, in our concrete case, it was given by the speed of the current sources and their control by the PC, and it is in the range of hundreds of microseconds. We did not define the goal of fast reconfiguration, as we were focused on testing and realising the concept of reconfigurable inverse-design magnonics. But the fundamental limitation, in this case, would be the inductance associated with the current loops, and a good reference to the limiting speed is the dynamic magnonic crystal studied in~\cite{chumak_all-linear_2010, chumak_current-controlled_2009}. There, they used a high inductance meander and realised the on/off of the field within about 5 to 10\,ns. This switching time was also defined by the power source rather than the device. Thus, we can claim that the reconfigurability of our developed specialised inverse-design device within a one-nanosecond timeframe is feasible, with its implementation being merely a technical matter.  Consequently, it possesses the capability to address the inverse problem in real time effectively.

We need to distinguish between two different timescales here. The first one is the "training time" the algorithm requires to learn and optimise a specific functionality provided by the user. In our experiments, this optimisation time can range from minutes to hours, depending on the problem complexity. However, once the optimised configuration has been determined, there is no need to repeat the optimisation process for the same functionality and the current configuration can be simply taken from a "library of saved configurations" optimised for different functionalities. The second timescale is the "reconfiguration time" needed to switch between different current configurations and consequently different functionalities. The switching time primarily depends on the speed of the current application and the inductance of the current loop, typically occurring on a nanosecond timescale.

\subsection{Perspectives on miniaturisation and non-volatility}
From an applied point of view, we do not consider the processor as a device that can compete with the surface acoustic wave (SAW) or bulk acoustic wave (BAW) microwave passive components such as filters and multiplexers currently used in mobile phones and other communication systems, but rather as a platform to study the performance and potential of the inverse design for 5G RF applications. For a competitive device, the miniaturization of the device is required both in the magnonic media (e.g. today we work with magnonic waveguides of 50\,nm width~\cite{heinz_propagation_2020}) as well as in the effective size of the introduced inhomogeneities. Moreover, the proposed proof-of-concept demonstrator uses considerable electrical currents and is energetically inefficient. Therefore, the formation of field inhomogeneities should be realised by other non-volatile methods well studied in the field of magnonics~\cite{barman_2021_2021,chumak_advances_2022}. The current loops could potentially be replaced by an array of individually controlled nanomagnets of a complex shape to achieve multiple degrees of freedom (simple-shaped nanomagnets were already used in~\cite{imre_majority_2006,kronast_element-specific_2011,haldar_reconfigurable_2016}). In this way, the RF device would consume energy only at the moment of reconfigurability of its operational parameters while remaining a passive device during operation. The feasibility of the experimental inverse design concept, however, is explicitly proven by the results presented here.

\section{Conclusion}\label{sec5}

We have developed a universal reconfigurable magnonic inverse-design device with a high degree of tunability, possessing 10$^{\text{87}}$ degrees of freedom. Through this work, we have demonstrated the device capabilities to successfully solve inverse-design problems experimentally, without the use of numerical simulations. Two functionalities of a reconfigurable RF notch filter and a demultiplexer were successfully realised on the same universal device. The reconfigurable notch filter had a 5\,MHz bandwidth for any desired center frequency, and the signal suppression within the BW increased by up to 48\,dB after the optimisation process was completed. The ability of the device to modulate the signal over more than four orders of magnitude highlights the exceptional performance and potential of the presented experimental inverse design concept. Similarly, the RF demultiplexer demonstrated high performance by efficiently guiding the two defined excited frequencies through the design region to be accurately detected at the desired outputs. The "training time" for any functionality can range from minutes to hours. However, once the optimised configuration is added to the saved "library of current configurations" for different functionalities, switching to any of them can occur on the nanosecond timescale.

Various aspects of experimental inverse design have been thoroughly explored in this study. First, we examined the device effective operational wavelength range, demonstrating that the device operating with wavelengths down to 70\,$\upmu$m, that are much smaller than the inhomogeneity dimension created by one omega loop, performs exceptionally well. Conversely, wavelengths larger than this dimension, up to 1.25\,mm, exhibit decreased efficiency due to the diffraction limit. Secondly, the study delves into the non-linear operational regime, highlighting that controllable non-stochastic spin-wave non-linearity enhances the device performance in finding better solutions to the inverse problem in shorter times. And lastly, two different algorithms, one of which is a machine-learning approach, have been employed successfully to optimise the device for multiple functionalities. 

This study successfully demonstrates linear functionalities tailored for RF communication systems, 5G, and future 6G technologies realised by the same reconfigurable universal inverse-design device. Leveraging an intricately designed reconfigurable region and employing intrinsic spin-wave non-linearity, our device can also achieve various nonlinear functionalities for logic gates, reservoir computing, and neuromorphic computing.

\section*{Acknowledgements}

The financial support by the Austrian Science Fund (FWF) via Grant No. I 4917-N (MagFunc) is acknowledged. A.C. acknowledges the financial support by the European Research Council (ERC) Proof of Concept Grant 101082020 5G-Spin. S.K. acknowledges the support by the H2020-MSCA-IF under Grant No. 101025758 ("OMNI"). Q.W. acknowledges the support from the National Key Research and Development Program of China (Grant No. 2023YFA1406600). We are grateful to Prof. Dr. D. Bozhko for his kind support in calculating the spin-wave dispersion curves in YIG film and to Prof. P. Pirro and Prof. G. Csaba for valuable discussions on the magnonic inverse-design device concept. We acknowledge the efforts of ElbaTech Srl in the development of the custom-made multichannel current sources.

\bibliography{ref}

\begin{thebibliography}{10}

\bibitem{noauthor_5g_nodate}
D.~Chandramouli, R.~Liebhart, J.~Pirskanen, {\it {5G} for the {Connected} {World}\/} (Wiley, 2019).

\bibitem{giribaldi_compact_2024}
G.~Giribaldi, L.~Colombo, P.~Simeoni, M.~Rinaldi, {\it Nature Communications\/} {\bf 15}, 304 (2024).

\bibitem{dieny_opportunities_2020}
B.~Dieny, {\it et~al.\/}, {\it Nature Electronics\/} {\bf 3}, 446 (2020).

\bibitem{gurevich_magnetization_1996}
A.~G. Gurevich, G.~A. Melkov, {\it Magnetization {Oscillations} and {Waves}\/} (CRC Press, 1996).

\bibitem{stancil_spin_2009}
D.~D. Stancil, A.~Prabhakar, {\it Spin {Waves}: {Theory} and {Applications}\/} (Springer Science \& Business Media, 2009).

\bibitem{v_v_kruglyak_magnonics_2010}
{V V Kruglyak}, {S O Demokritov}, {D Grundler}, {\it Journal of Physics D: Applied Physics\/} {\bf 43}, 260301 (2010).

\bibitem{barman_2021_2021}
A.~Barman, {\it et~al.\/}, {\it Journal of Physics: Condensed Matter\/} {\bf 33}, 413001 (2021).

\bibitem{chumak_advances_2022}
A.~V. Chumak, {\it et~al.\/}, {\it IEEE Transactions on Magnetics\/} {\bf 58}, 1 (2022).

\bibitem{wu_high-performance_2017}
Y.~Wu, {\it et~al.\/}, {\it Advanced Materials\/} {\bf 29}, 1603031 (2017).

\bibitem{wang_nanoscale_2024}
Q.~Wang, G.~Csaba, R.~Verba, A.~V. Chumak, P.~Pirro, {\it Physical Review Applied\/} {\bf 21}, 040503 (2024).

\bibitem{wang_magnonic_2020}
Q.~Wang, {\it et~al.\/}, {\it Nature Electronics\/} {\bf 3}, 765 (2020).

\bibitem{wu_magnon_2018}
H.~Wu, {\it et~al.\/}, {\it Physical Review Letters\/} {\bf 120}, 097205 (2018). Publisher: American Physical Society.

\bibitem{chumak_magnon_2014}
A.~V. Chumak, A.~A. Serga, B.~Hillebrands, {\it Nature Communications\/} {\bf 5}, 4700 (2014).

\bibitem{mahmoud_introduction_2020}
A.~Mahmoud, {\it et~al.\/}, {\it Journal of Applied Physics\/} {\bf 128}, 161101 (2020).

\bibitem{torrejon_neuromorphic_2017}
J.~Torrejon, {\it et~al.\/}, {\it Nature\/} {\bf 547}, 428 (2017).

\bibitem{papp_nanoscale_2017}
A.~Papp, W.~Porod, Ã.~I. Csurgay, G.~Csaba, {\it Scientific Reports\/} {\bf 7}, 9245 (2017).

\bibitem{bracher_analog_2018}
T.~Brächer, P.~Pirro, {\it Journal of Applied Physics\/} {\bf 124}, 152119 (2018).

\bibitem{wang_inverse-design_2021}
Q.~Wang, A.~V. Chumak, P.~Pirro, {\it Nature Communications\/} {\bf 12}, 2636 (2021).

\bibitem{kiechle_experimental_2022}
M.~Kiechle, {\it et~al.\/}, {\it IEEE Magnetics Letters\/} {\bf 13}, 1 (2022).

\bibitem{papp_nanoscale_2021}
A.~Papp, W.~Porod, G.~Csaba, {\it Nature Communications\/} {\bf 12}, 6422 (2021).

\bibitem{chumak_all-linear_2010}
A.~V. Chumak, {\it et~al.\/}, {\it Nature Communications\/} {\bf 1}, 141 (2010).

\bibitem{bozhko_unconventional_2020}
D.~A. Bozhko, {\it et~al.\/}, {\it Physical Review Research\/} {\bf 2}, 023324 (2020).

\bibitem{dubs_sub-micrometer_2017}
C.~Dubs, {\it et~al.\/}, {\it Journal of Physics D: Applied Physics\/} {\bf 50}, 204005 (2017).

\bibitem{serga_yig_2010}
A.~A. Serga, A.~V. Chumak, B.~Hillebrands, {\it Journal of Physics D: Applied Physics\/} {\bf 43}, 264002 (2010).

\bibitem{abert_magnumfe_2013}
C.~Abert, L.~Exl, F.~Bruckner, A.~Drews, D.~Suess, {\it Journal of Magnetism and Magnetic Materials\/} {\bf 345}, 29 (2013).

\bibitem{abert_micromagnetics_2019}
C.~Abert, {\it The European Physical Journal B\/} {\bf 92}, 120 (2019).

\bibitem{schrefl_numerical_2007}
T.~Schrefl, {\it et~al.\/}, {\it Numerical {Methods} in {Micromagnetics} ({Finite} {Element} {Method})\/} (Wiley, 2007).

\bibitem{wang_deeply_2023}
Q.~Wang, {\it et~al.\/}, {\it Science Advances\/} {\bf 9}, eadg4609 (2023).

\bibitem{connelly_efficient_2021}
D.~A. Connelly, {\it et~al.\/}, {\it Scientific Reports\/} {\bf 11}, 18378 (2021).

\bibitem{kalinikos_theory_1986}
B.~A. Kalinikos, A.~N. Slavin, {\it Journal of Physics C: Solid State Physics\/} {\bf 19}, 7013 (1986).

\bibitem{serga_parametrically_2007}
A.~A. Serga, {\it et~al.\/}, {\it Physical Review Letters\/} {\bf 99}, 227202 (2007).

\bibitem{vilsmeier_spatial_2024}
F.~Vilsmeier, C.~Riedel, C.~H. Back, {\it Applied Physics Letters\/} {\bf 124}, 132407 (2024).

\bibitem{gad_pygad_2023}
A.~F. Gad, {\it Multimedia Tools and Applications\/}  (2023).

\bibitem{shen_integrated-nanophotonics_2015}
B.~Shen, P.~Wang, R.~Polson, R.~Menon, {\it Nature Photonics\/} {\bf 9}, 378 (2015).

\bibitem{molesky_inverse_2018}
S.~Molesky, {\it et~al.\/}, {\it Nature Photonics\/} {\bf 12}, 659 (2018).

\bibitem{hughes_adjoint_2018}
T.~Hughes, M.~Minkov, I.~Williamson, S.~Fan, {\it ACS Photonics\/} {\bf 5}, 4781 (2018).

\bibitem{wang_reconfigurable_2018}
Q.~Wang, {\it et~al.\/}, {\it Science Advances\/} {\bf 4}, e1701517 (2018).

\bibitem{ge_nanoscaled_2024}
X.~Ge, R.~Verba, P.~Pirro, A.~V. Chumak, Q.~Wang, {\it Applied Physics Letters\/} {\bf 124}, 122413 (2024).

\bibitem{chumak_current-controlled_2009}
A.~V. Chumak, T.~Neumann, A.~A. Serga, B.~Hillebrands, M.~P. Kostylev, {\it Journal of Physics D: Applied Physics\/} {\bf 42}, 205005 (2009).

\bibitem{heinz_propagation_2020}
B.~Heinz, {\it et~al.\/}, {\it Nano Letters\/} {\bf 20}, 4220 (2020).

\bibitem{imre_majority_2006}
A.~Imre, {\it et~al.\/}, {\it Science\/} {\bf 311}, 205 (2006).

\bibitem{kronast_element-specific_2011}
F.~Kronast, {\it et~al.\/}, {\it Nano Letters\/} {\bf 11}, 1710 (2011).

\bibitem{haldar_reconfigurable_2016}
A.~Haldar, D.~Kumar, A.~O. Adeyeye, {\it Nature Nanotechnology\/} {\bf 11}, 437 (2016).

\end{thebibliography}

\bibliographystyle{Science}

\end{document}